# Anatomically and metabolically informed diffusion for unified denoising and segmentation in low-count PET imaging

Menghua Xia [a,b], Kuan-Yin Ko [c], Der-Shiun Wang [d,e], Ming-Kai Chen [a,b], Qiong Liu [f], Huidong Xie [f], Liang Guo [f], Wei Ji [a], Jinsong Ouyang [a,b], Reimund Bayerlein [g,h], Benjamin A. Spencer [h], Quanzheng Li [i], Ramsey D. Badawi [g,h], Georges El Fakhri [a,b], Chi Liu [a,b,f,*]

[a] Department of Radiology and Biomedical Imaging, Yale University School of Medicine, New Haven, CT, USA
[b] Yale Biomedical Imaging Institute, Yale University, New Haven, CT, USA
[c] Department of Nuclear Medicine, National Taiwan University Cancer Center, Taipei, Taiwan
[d] Department of Pediatric, Tri-service General Hospital, Taipei, Taiwan
[e] Department of Pediatric, School of Medicine, National Defenese Medical Center, Taipei, Taiwan
[f] Department of Biomedical Engineering, Yale University, New Haven, CT, USA
[g] Department of Biomedical Engineering, University of California Davis, Sacramento, CA, USA
[h] Department of Radiology, University of California Davis, Sacramento, CA, USA
[i] Department of Radiology, Massachusetts General Hospital and Harvard Medical School, Boston, MA, USA

ARTICLE INFO

ABSTRACT

Positron emission tomography (PET) image denoising, along with lesion and organ segmentation, are critical steps in PET-aided diagnosis. However, existing methods typically treat these tasks independently, overlooking inherent synergies between them as correlated steps in the analysis pipeline. In this work, we present the anatomically and metabolically informed diffusion (AMDiff) model, a unified framework for denoising and lesion/organ segmentation in low-count PET imaging. By integrating multi-task functionality and exploiting the mutual benefits of these tasks, AMDiff enables direct quantification of clinical metrics, such as total lesion glycolysis (TLG), from low-count inputs. The AMDiff model incorporates a semantic-informed denoiser based on diffusion strategy and a denoising-informed segmenter utilizing nnMamba architecture. The segmenter constrains denoised outputs via a lesion-organ-specific regularizer, while the denoiser enhances the segmenter by providing enriched image information through a denoising revision module. These components are connected via a warming-up mechanism to optimize multi-task interactions. Experiments on multi-vendor, multi-center, and multi-noise-level datasets demonstrate the superior performance of AMDiff. For test cases below 20% of the clinical count levels from participating sites, AMDiff achieves TLG quantification biases of −21.60±47.26%, outperforming its ablated versions which yield biases of −30.83±59.11% (without the lesion-organ-specific regularizer) and −35.63±54.08% (without the denoising revision module). By leveraging its internal multi-task synergies, AMDiff surpasses standalone PET denoising and segmentation methods. Compared to the benchmark denoising diffusion model, AMDiff reduces the normalized root-mean-square error for lesion/liver by 22.92/17.27% on average. Compared to the benchmark nnMamba segmentation model, AMDiff improves lesion/liver Dice coefficients by 10.17/2.02% on average.

## 1. Introduction

Positron emission tomography (PET) is a highly sensitive nuclear medicine imaging technique widely employed in oncology, neurology, and cardiology (Kitson et al., 2009). Reconstructed PET images, usually complemented by computed tomography (CT) images, are used to identify lesions, locate organs, and quantify clinical metrics such as metabolic tumor volume (MTV), standardized uptake value (SUV) within regions of interest (ROIs), and total lesion glycolysis (TLG), thereby aiding disease diagnosis and treatment planning (Chen et al., 2012), as illustrated in Fig. 1. Image quality and precision of downstream semantic analysis are two of the critical factors that influence accurate quantification. Image quality is primarily determined by noise level, which depends on the number of detected photon events during PET acquisition. These photon counts, in turn, depend on factors such as radiotracer dose, scan duration, and scanner efficiency. Precise semantic analysis demands meticulous annotation by experienced physicians, a process that is both labor-intensive and time-consuming. Recently, deep learning has shown great potential in addressing these challenges through image denoising and automatic segmentation. Denoising techniques enable the generation of high-count PET images from low-count acquisitions (Bousse et al., 2024), allowing for reduced radiotracer doses and shorter scan times, thus enhancing patient comfort and reducing motion artifacts. Meanwhile, automatic segmentation can significantly reduce physician workload by efficiently delineating ROIs (Yousefirizi et al., 2021).

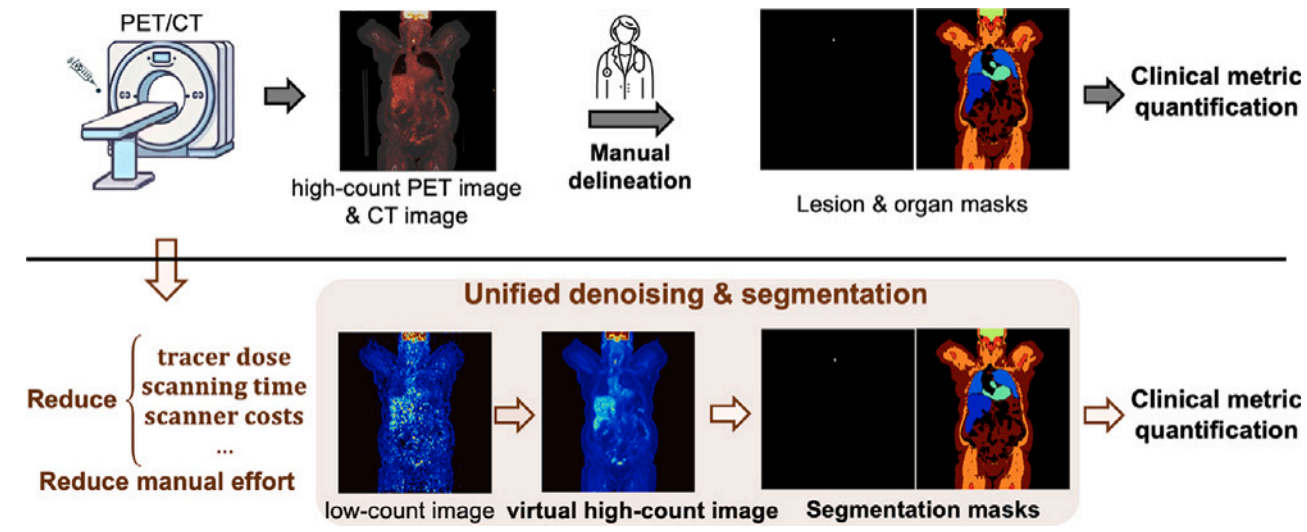

**Fig. 1.** Deep learning PET image analysis offers substantial benefits. This work investigates unified denoising and lesion and organ segmentation in low-count PET imaging while leveraging the synergies between these tasks.

Deep learning PET image denoising has progressed significantly with the development of models covering convolutional neural networks (CNNs) (Angelis et al., 2021; Liu et al., 2022), generative adversarial networks (GANs) (Xue et al., 2022; Zhu et al., 2023), vision transformers (ViTs) (Jang et al., 2023), and, more recently, denoising diffusion probabilistic models (DDPMs) (Han et al., 2023; Gong et al., 2024a; Xie et al., 2024). Additionally, some models incorporate functionalities to enhance practical applicability, such as noise-level adaptation mechanisms (Xie et al., 2023) and cross-center data privacy considerations (Zhou et al., 2023). Despite these advancements, deep learning-denoised PET images still face critical challenges, including over-smoothing that reduces lesion contrast and detectability, as well as hallucinations where artificial features, such as non-existent lesions, are introduced (Xia et al., 2025). These limitations compromise the anatomical and metabolic fidelity of the denoised images, restricting their utility for downstream clinical analysis. To address these issues, several approaches incorporate auxiliary semantic priors as regularization mechanisms. For instance, semantic features extracted via additional convolutional layers from co-registered MR or CT images are integrated into the PET denoising pipeline through feature concatenation or attention mechanisms (Fu et al., 2024; Onishi et al., 2021; Cui et al., 2019). Other studies leverage priors derived from the original sinogram data to guide the denoising process (Zhang et al., 2024). These methods rely on intermediate modules external to the core denoising task and extract indirect or implicit semantic cues from multi-modal inputs. A more direct strategy may involve using explicit semantic labels, such as organ and tumor annotations, to supervise the denoising process. For example, (Huang et al., 2022; Xia et al., 2024) incorporate model-generated semantic labels into the loss function to regularize training. However, the accuracy of these generated labels may be suboptimal, potentially introducing bias into the denoising process.

Deep learning PET image segmentation primarily utilizes Unet-based architectures and has been explored for both lesion (Leung et al., 2024; Gatidis et al., 2024) and organ segmentation (Shiyam Sundar et al., 2022; Suganuma et al., 2023). For lesion segmentation, semi-supervised transfer learning is commonly used (Leung et al., 2024), leveraging large publicly available datasets (Gatidis et al., 2022) with full annotations for pre-training, followed by fine-tuning on local datasets with incomplete annotations. More recently, DDPMs have been employed to generate lesion-absent images from lesion-present ones, facilitating lesion detection through image subtraction (Ahamed et al., 2024). DDPMs have also been used to synthesize lesion-present images (Hu et al., 2024), serving as data augmentation for training segmentation networks. For organ segmentation, while most models (Shiyam Sundar et al., 2022; Suganuma et al., 2023) rely on PET-CT pairs as input, recent studies have shown the feasibility of multi-organ segmentation on CT-free PET scans (Liebgott et al., 2021; Salimi et al., 2022). These advancements are enabled by the growing capabilities of deep learning on larger datasets. Yet, all the aforementioned PET segmentation models, whether for lesions or organs, depend on standard- or high-count PET inputs to achieve satisfactory results. To the best of our knowledge, none has proven effective on low-count noisy inputs.

Denoising and segmentation, as naturally correlated tasks within the analysis workflow (illustrated in Fig. 1), logically exhibit synergistic interactions. Higher-quality denoised images, with enhanced visibility and differentiation of lesions and organs, simplify semantic segmentation. Vice versa, downstream lesion and organ masks can regularize the denoised outputs in terms of semantic structures, thereby supporting the denoising in a direct and explicit manner.

The concept of this multi-task learning has been actively explored across various domains in recent years, though it has yet to be specifically applied to PET denoising and segmentation. Multi-task learning has shown superior performance over traditional single-task learning by facilitating information sharing across tasks. For example, a multi-task diffusion framework (Ye and Xu, 2024) has been developed to jointly address multiple scene-related dense prediction tasks, such as segmentation and image generation. This framework leverages a conditioning mechanism that incorporates initial predictions from multiple auxiliary decoders to enhance the learning process of a target task. Similarly, image translation models have been designed with semantic or class label guidance (Peng et al., 2023; Li et al., 2023; Xu et al., 2023; Lim, 2023), where segmentation or classification models supervise the translation process, and the improved translations, in turn, enhance segmentation or classification performance. Despite their success in respective domains, these approaches face limitations when applied to PET analysis. They are primarily designed for processing 2D images and are not well-suited for handling 3D medical volume data (Ye and Xu, 2024; Li et al., 2023; Lim, 2023). Furthermore, auxiliary-task training for generating priors for the main task is often conducted independently of the main-task training (Ye and Xu, 2024; Peng et al., 2023), potentially introducing biases during the prior generation process. Alternatively, iterative training approaches may be adopted to mitigate biases (Xu et al., 2023), but this significantly increases training complexity.

In summary, from a clinical perspective, although numerous methods have been proposed for low-count PET denoising and standard-count PET segmentation, these two closely related tasks have typically been treated in isolation. As a result, existing approaches are unable to automatically derive clinically important metrics, such as TLG, directly from low-count inputs. A unified model that performs simultaneous denoising and segmentation, as illustrated in Fig. 1, offers a promising solution to this limitation. Such a unified model has the potential to reduce both scanning costs and physician workload through automated end-to-end processing. From a technical perspective, while mutual benefits between PET denoising and segmentation are anticipated, due to their shared semantic features, they remain largely underexplored. Existing multi-task learning models designed for other imaging tasks often encounter challenges when applied to PET, mainly due to mismatches in data dimensionality and limitations in cross-task interaction design.

In this work, we propose an innovative multi-task framework for unified denoising and segmentation for low-count PET imaging, named Anatomically and Metabolically Informed Diffusion (AMDiff). Our contributions are threefold. **(1)** The AMDiff enables one-step clinical metric quantification directly from low-count inputs by simultaneously and automatically generating denoised images along with lesion and organ masks. **(2)** The AMDiff fully explores synergies between tasks, with segmentation guiding denoising on semantic constructions and denoising facilitating more robust segmentation. The two tasks are trained jointly, with direct access to inputs and labels, and are interconnected through a warming-up mechanism that facilitates efficient and accurate information exchange. **(3)** We demonstrate the effectiveness of AMDiff compared to state-of-the-art (SOTA) denoising and segmentation methods, through comprehensive experiments conducted on datasets from multiple vendors, centers, and noise levels.

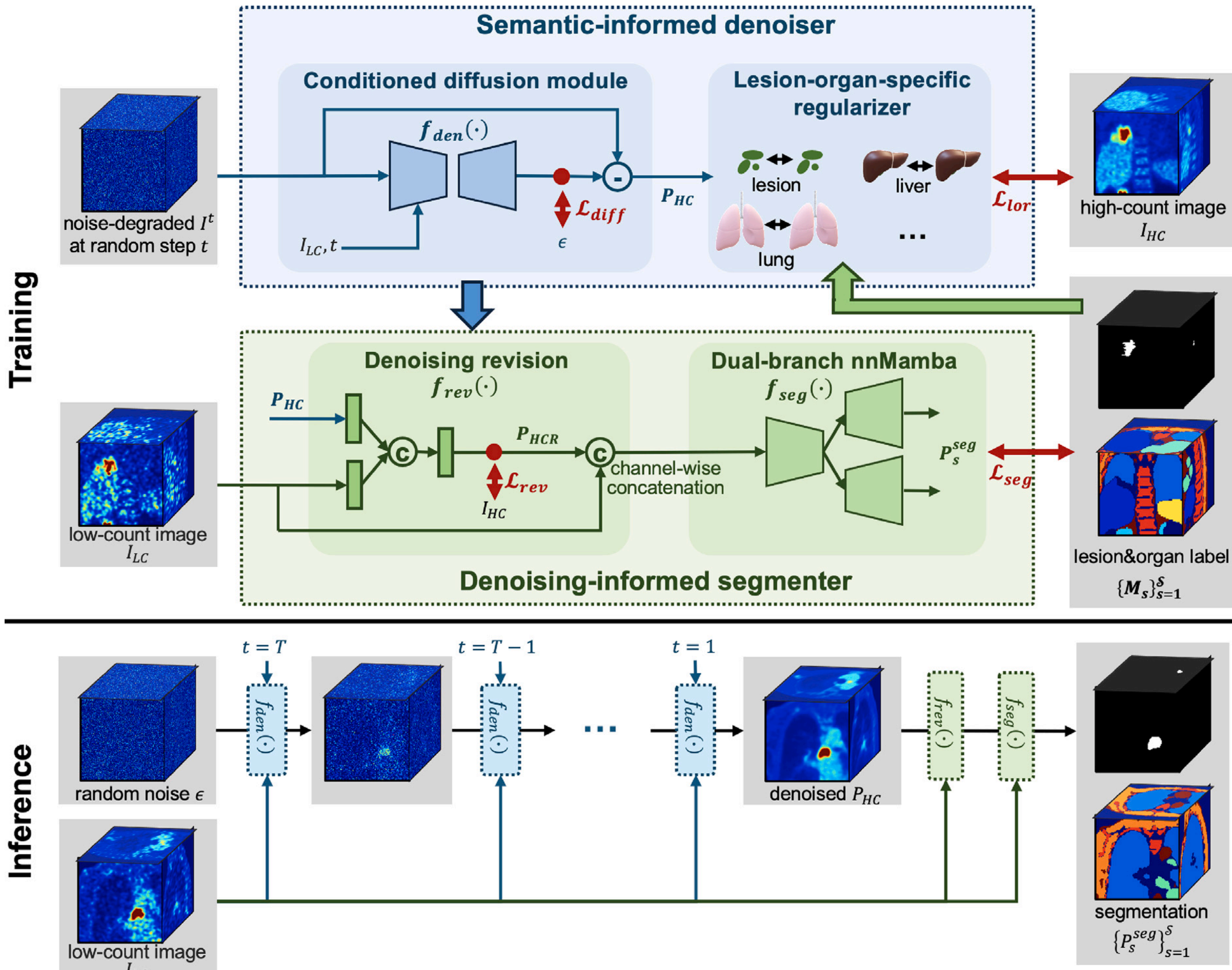

**Fig. 2.** Overview of the AMDiff model. It comprises a semantic-informed denoiser and a denoising-informed segmenter. The segmenter constrains the semantic structures of denoised outputs using a lesion-organ-specific regularizer. Vice versa, the denoiser supports the segmenter by providing images with enhanced lesion visibility and organ clarity via the denoising revision module.

## 2. Method

The framework of AMDiff is shown in Fig. 2. It comprises a semantic-informed denoiser and a denoising-informed segmenter. The denoiser reconstructs high-count images $I_{HC}$ from low-count ones $I_{LC}$. The denoising process is constrained by a lesion-organ-specific regularizer, which emphasizes anatomical and metabolic similarities of denoised outputs $P_{HC}$ with references $I_{HC}$. The segmenter inherits denoised information $P_{HC}$ via a revision function $f_{rev}(\cdot)$ and generates complete lesion and organ masks using a dual-branch architecture $f_{seg}(\cdot)$. Due to computational constraints, large PET images are divided into smaller 3D patches for network processing and are reassembled afterward. Details of the AMDiff components are as below.

### 2.1. *Semantic-informed denoiser*

To ensure reliability for clinical analysis, the denoised output $P_{HC}$ must align with the high-count reference $I_{HC}$ at the level of each lesion and organ. To achieve this, the semantic-informed denoiser incorporates a conditioned diffusion module with the denoising function $f_{den}(\cdot)$ as its baseline, optimized under the guidance of a novel lesion-organ-specific regularizer $\mathcal{L}_{lor}$.

#### 2.1.1. *Conditioned diffusion module*

In the forward diffusion process, the high-count PET image $I_{HC}$ is progressively degraded by adding Gaussian noise step-by-step along a Markovian chain with $T$ total steps, as illustrated in Fig. 2. Due to the additive property of the Gaussian distribution, the noise-degraded image at any step $t$ can be expressed as (Ho et al., 2020):

$$I^t = \sqrt{\bar{\alpha}_t} I_{HC} + \sqrt{1 - \bar{\alpha}_t}\,\epsilon, \quad 1 \le t \le T \tag{1}$$

where $\{\alpha_t \mid 1 \le t \le T\}$ represent hyperparameters of a predefined variance scheduler. $\bar{\alpha}_t = \prod_{k=1}^{t} \alpha_k$ denotes the cumulative product of $\alpha$ from 1 to $t$. The term $\epsilon \sim \mathcal{N}(0, \mathbf{I})$ is a noise map randomly sampled from the standard Gaussian distribution.

A denoising function $f_{den}(\cdot)$ is trained to predict the added noise $\epsilon$, conditioned on the low-count image $I_{LC}$ and optimized using the loss function $\mathcal{L}_{diff}$ (Dorjsembe et al., 2024), formulated as:

$$\epsilon_{recon} = f_{den}(I^t \copyright I_{LC}, t) \tag{2}$$

$$\mathcal{L}_{diff} = \|\epsilon_{recon} - \epsilon\|_1 \tag{3}$$

where © represents channel-wise concatenation. The function $f_{den}(\cdot)$ is implemented using a Unet architecture, consisting of a four-layer encoder and a three-layer decoder, with self-attention incorporated in the lowest-resolution layer.

In the reverse diffusion process during inference, the trained denoiser $f_{den}(\cdot)$ reverses the forward diffusion trajectory, starting from Gaussian noise and iteratively generating $\{I^{T-1}, I^{T-2}, \ldots, I^0\}$ (Ho et al., 2020) with the low-count $I_{LC}$ as a condition (Eq. (4)), as shown in Fig. 2. The final output $I^0$ becomes the denoised result $P_{HC}$, approximating the high-count image.

$$I^{t-1} = \frac{1}{\sqrt{\alpha_t}}\left(I^t - \frac{1-\alpha_t}{\sqrt{1-\bar{\alpha}_t}} \cdot f_{den}(I^t \circledcirc I_{LC}, t)\right) + \sigma_t z \tag{4}$$

The term $z \sim \mathcal{N}(0, \mathbf{I})$.

#### 2.1.2. Lesion-organ-specific regularizer

The lesion-organ-specific regularizer is designed to ensure that the denoised prediction $P_{HC}$ corresponds to the high-count reference $I_{HC}$ in terms of semantic structures. To achieve this, it employs a loss function $\mathcal{L}_{lor}$, which utilizes lesion and organ labels to constrain $P_{HC}$, formulated as:

$$\mathcal{L}_{lor} = \sum_{s=1}^{S} w_s \cdot \|M_s \cdot (P_{HC} - I_{HC})\|_1 \tag{5}$$

where $M_s$ and $w_s$ represent the binary mask and assigned weight for the semantic class $s$ ($1 \leq s \leq S$), respectively. $S$ denotes the total number of segmentation classes. Specifically, $s = 1$ corresponds to the lesion class, while $s = \{2, \ldots, S\}$ corresponds to the organ classes.

### 2.2. Denoising-informed segmenter

To enable reliable segmentation on low-count inputs, the denoising-informed segmenter is designed to utilize the obtained denoised information $P_{HC}$ as auxiliary input. The segmenter first applies a revision function $f_{rev}(\cdot)$ to forward the denoised information, followed by lesion and organ segmentation using a dual-branch nnMamba (Gong et al., 2024b) architecture $f_{seg}(\cdot)$.

#### 2.2.1. Denoising revision

The denoising revision module $f_{rev}(\cdot)$ serves to bridge the denoiser and segmenter while recovering the full SUV data range. In PET images, SUV values span a wide range from 0 to 100, occasionally exceeding 200 in some cases, though most diagnostically relevant values are concentrated within the 0–20 range. To simplify diffusion training, a cutoff is applied to SUV values above 20. Subsequently, the revision module $f_{rev}(\cdot)$ maps the diffusion output $P_{HC}$ to a revised map $P_{HCR}$ covering the full data range. Within $f_{rev}(\cdot)$, a direct skip connection to the original low-count input $I_{LC}$ is included to restore information lost due to the data range cutoff, as illustrated in Fig. 2. $P_{HC}$ and $I_{LC}$ are passed through consecutive convolutional layers to produce the revised map $P_{HCR}$. Intermediate deep supervision, $\mathcal{L}_{rev}$, is applied to $P_{HCR}$ using a class-weighted format similar to Eq. (5), formulated as:

$$P_{HCR} = f_{rev}(P_{HC}, I_{LC}) \tag{6}$$

$$\mathcal{L}_{rev} = \sum_{s=0}^{S} w_s \cdot \|M_s \cdot (P_{HCR} - I_{HC})\|_1 \tag{7}$$

where $s = 0$ corresponds to the background.

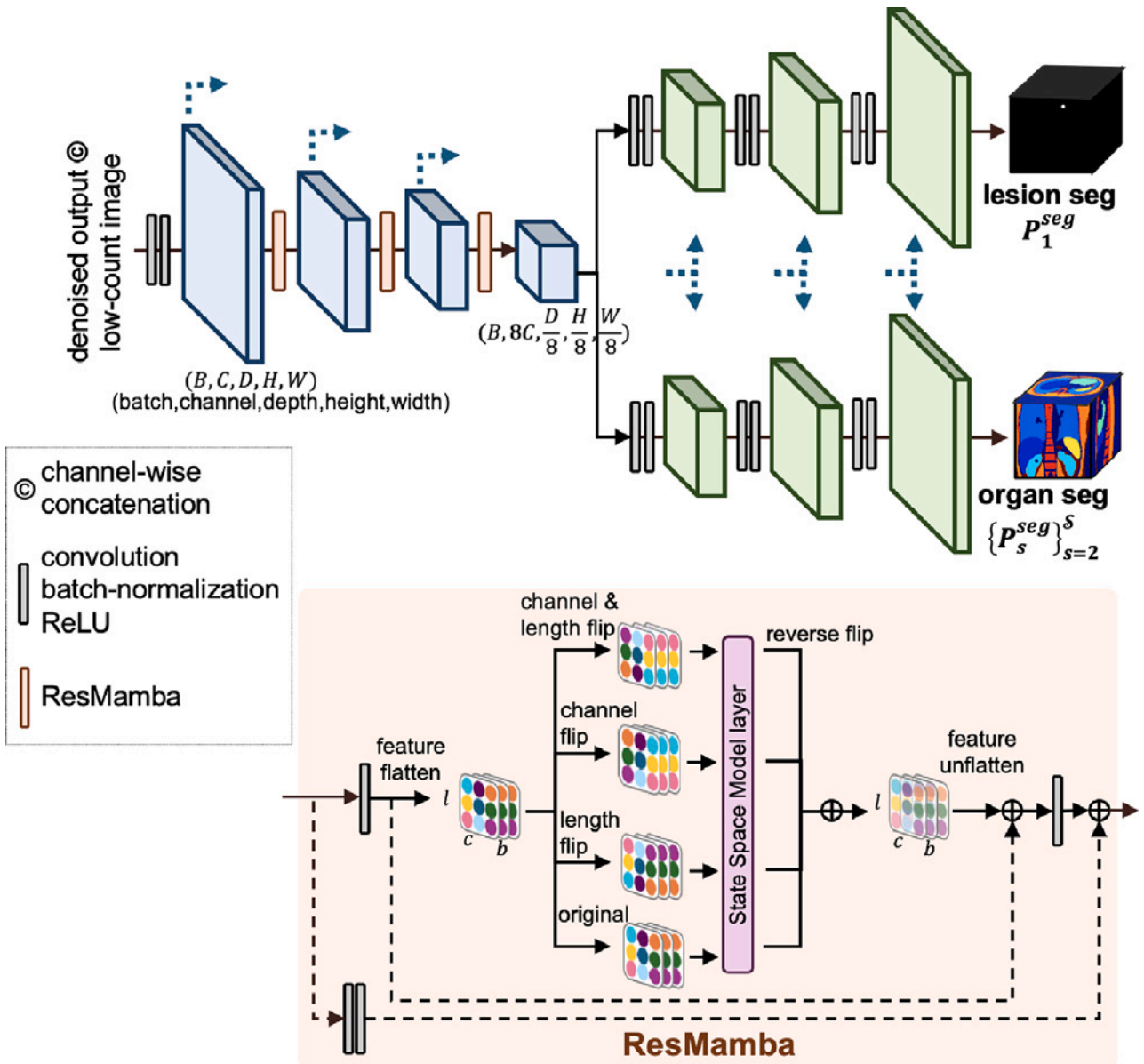

**Fig. 3.** The dual-branch segmenter architecture of the AMDiff model.

#### 2.2.2. Dual-branch nnMamba

The segmentation module $f_{seg}(\cdot)$ generates both lesion and organ masks from a two-channel input, which concatenates the revised $P_{HCR}$ and the original low-count $I_{LC}$. The nnMamba architecture is adopted for its ability to combine the strengths of convolutional layers in local pattern modeling with Mamba layers in long-range dependency modeling (Gong et al., 2024b). Since lesion and organ segmentation are closely related tasks sharing similar visual features, a dual-branch architecture with a shared encoder and separate decoders is designed to minimize overall model complexity. As depicted in Fig. 3, the encoder extracts visual features through an initial convolutional layer, which are then processed through three consecutive Mamba-in-convolution layers (ResMamba) to capture both local and global contextual information. In ResMamba, features are augmented across channel and spatial dimensions before being processed by a state-space model layer, maximizing their representational capacity. The encoded features are then forwarded to two separate three-layer decoders, generating the segmentation masks.

The segmentation loss combines cross-entropy loss and focal Dice loss, formulated as the first and second terms in Eq. (9), respectively. The inclusion of focal loss helps mitigate the class imbalance problem to some extent (Yeung et al., 2022).

$$P^{seg} = f_{seg}(P_{HCR} \circledcirc I_{LC}) \tag{8}$$

$$\mathcal{L}_{seg} = \frac{1}{S}\sum_{s=1}^{S} w_s \Bigg(-M_s \log(P_s^{seg}) \quad + \left[1 - \frac{2 \cdot (1-P_s^{seg})^{4/3} \cdot P_s^{seg} \cdot M_s}{(1-P_s^{seg})^{4/3} \cdot (P_s^{seg})^2 + (M_s)^2}\right]\Bigg) \tag{9}$$

$M_s$ and $P_s^{seg}$ represent the binary label mask and the predicted probability for the semantic class $s$, respectively. $w_s$ denotes the weight assigned to each class.

### 2.3. Unified multi-task learning

The AMDiff was trained as a whole, with the total optimization loss function expressed as:

$$\mathcal{L}_{total} = \mathcal{L}_{diff} + \lambda_{warm}(\mathcal{L}_{lor} + \mathcal{L}_{rev} + \mathcal{L}_{seg}) \tag{10}$$

$$\lambda_{warm} = (e \geq e_T) \cdot \exp(-5 \cdot (1 - e/e_{max})^2) \quad (11)$$

where $\lambda_{warm}$ denotes an epoch-dependent warm-up weight. At the initial stage of training, only the denoiser is optimized during the first $e_T$ epochs, as the denoised outputs are not yet sufficiently reliable to guide the segmentation task. As training progresses and the quality of the denoised information improves, the segmenter is gradually incorporated into the training process. The weight $\lambda_{warm}$ increases progressively, allowing the segmenter to benefit from increasingly accurate inputs and enabling effective synergy between denoising and segmentation. In this formulation, $e$ denotes the current training epoch, and $e_{max}$ is the total number of training epochs.

## 3. Experiments

### 3.1. Materials

We evaluated the AMDiff model on multi-vendor and multi-center $^{18}$F-FDG PET datasets, as detailed in Table 1. **Dataset A** was acquired at Yale University, USA, using a Siemens Biograph mCT scanner and included two groups. Group I consisted of 195 subjects scanned with a single pass of continuous bed motion (CBM) acquisition starting from 60 min post-injection, generating standard- or normal-count images through listmode data rebinning. Group II included 29 subjects scanned using 19-pass CBM scanning over 90 min, creating high-count images by combining list-mode data from all passes after 60 min post-injection (Liu et al., 2022). Low-count images were created by non-overlapping down-sampling the PET list-mode data. Subjects in Group I were used for training and validation, while all subjects in Group II were reserved for testing. **Dataset B** was also collected at Yale University, and included 10 subjects scanned on a Siemens Biograph Vision scanner. High-count images were generated similarly using 19-pass CBM scanning over 90 min, while standard-count images were generated by a single 5-min pass. All subjects in Dataset B were used exclusively for testing, with models trained on Group I from Dataset A. This setup was designed to evaluate the model's generalizability to data from a scanner not seen during training. **Dataset C** was obtained from UC Davis Medical Center, USA, and included 10 subjects scanned on a United Imaging uExplorer scanner. These scans were conducted for 20 min starting from 120 min post-injection. Low-count images were created by non-overlapping down-sampling the PET list-mode data. The model was trained and evaluated independently on Dataset C, using three subjects for training, two for validation, and five for testing.

A set of lesion labels was collaboratively delineated by two physicians. Organ labels were generated using TotalSegmentator (Wasserthal et al., 2023), a CT-based organ segmentation tool, applied to PET-paired CT images. Obvious segmentation errors were subsequently corrected through manual annotation. While the tool provided labels for over 100 organ classes, this study focused on classes of high interest: liver, lung, bone, muscle, kidney, spleen, and aorta.

### 3.2. Technical details and evaluation metrics

The patch size for dividing images during network processing was set to $128 \times 128 \times 128$ voxels, with a stride size of $32 \times 32 \times 32$ voxels. The adaptive patch sampling strategy in Xia et al. (2024) was employed to balance lesion-present and lesion-absent samples. Diffusion training used a cosine variance scheduler, defining the hyperparameters $\alpha_t$ within the range (0,1) and setting the total timesteps to $T = 250$.

The AMDiff model was implemented on the PyTorch platform and deployed on a Nvidia H100 GPU. Training was conducted $e_{max}$=800 epochs using the Adam optimizer, with an initial learning rate of $10^{-5}$ for the denoiser and $10^{-3}$ for the segmenter during the first 400 epochs, which was subsequently reduced by a factor of 0.1 for the remaining epochs. The parameter $e_T$ in Eq. (11) was set to 100, meaning that only the denoiser was trained during the first 100 epochs to stabilize the initial training stage.

For denoising performance evaluations, the normalized root-mean-square error (NRMSE, normalized by the range of observed data) was computed between denoised results and high-count references. This metric was assessed for the entire 3D image and specific classes, including lesion, liver, lung, bone, muscle, kidney, spleen, and aorta. For segmentation performance evaluations, the widely used Dice coefficient was adopted to measure performance for both lesion and organ classes. Finally, to evaluate the unified model as a whole, clinical metrics including the MTV, TLG, and organ $SUV_{mean}$ were automatically computed by the AMDiff model from low-count inputs, and were compared against ground truth values derived from high-count images with manual annotations.

## 4. Results

### 4.1. Denoising evaluations

The denoising performance of AMDiff was compared against representative denoising methods, including LeqMod-GAN (Xia et al., 2024), SpachTransformer (Jang et al., 2023), and Med-DDPM (Dorjsembe et al., 2024). All comparison models were re-trained on the local datasets using their official implementations, with whole images divided into 3D patches of 128 × 128 × 128 voxels. For each method, the version that achieved the best performance on the validation set was selected for final testing.

Results are summarized in Table 2. Unlike the comparison models, which are limited to denoising functionality, the AMDiff integrates segmentation maps to regularize denoised outputs, demonstrating superior performance across multiple scanners. When averaged across all test cases from all scanners and noise levels, AMDiff achieved liver NRMSE reductions of 31.6%, 7.1%, and 17.3% compared to LeqMod, SpachTransformer, and Med-DDPM, respectively. For lesion regions, the corresponding reductions were 2.6%, 8.9%, and 22.9%. In particularly challenging cases with extremely low-count levels below 5%, AMDiff delivered even greater improvements in lesion NRMSE, achieving reductions of 12.0%, 24.1%, and 32.5%.

For visual comparisons, Fig. 4 presents sample denoised images generated by different models. Relative to high-count references, methods such as LeqMod-GAN, SpachTransformer, and Med-DDPM exhibit reduced lesion contrast and blurred anatomical boundaries, particularly for small lesions and bone structures. In contrast, AMDiff leverages both anatomical and metabolic guidance to produce semantically coherent reconstructions that closely resemble high-count references, as highlighted by red arrows in Fig. 4. In certain cases, lesion characteristics in AMDiff-denoised images differ from those in high-count references and may appear as hallucinations, as indicated by the yellow arrows. This is likely because similar features exhibit strong signals in the low-count inputs, causing the denoised images to preserve these characteristics. Nevertheless, these results remain superior to those produced by other comparison models. There are also instances where small, faint lesions are missed by all denoising methods, likely due to their extremely low signal and the challenge of distinguishing them from high noise in low-count inputs, as indicated by the orange arrows in Fig. 4.

### 4.2. Segmentation evaluations

The segmentation performance of AMDiff was compared with SOTA segmentation methods, including SwinUNETR (He et al., 2023), nnMamba (Gong et al., 2024b), and Med-SAM (Wang et al., 2024). To ensure a fair comparison, the input to each comparison model was the concatenation of the low-count images and the denoised outputs generated by the same conditioned diffusion module architecture used in AMDiff.

**Table 1**
Details of used datasets.

| | DatasetA | | DatasetB | DatasetC |
|---|---|---|---|---|
| Medical center | Yale University | | | UC Davis |
| Scanner | Siemens Biograph mCT | | Siemens Biograph Vision | United Imaging uExplorer |
| | Group I | Group II | | |
| Average dose (mean ± std,MBq) | 256.3 ± 16.2 | 334.5 ± 32.1 | 332.1 ± 28.0 | 300.6 ± 9.0 |
| OSEM parameters | 2 iterations, 21 subsets | 2 iterations, 21 subsets | 2 iterations, 5 subsets | 4 iterations, 20 subsets |
| FWHM in Gaussian smoothing | 5 mm | 5 mm | 2 mm | no Gaussian filtering |
| Image size (voxels) | $440 \times 440 \times h$[a] | $440 \times 440 \times h$[a] | $440 \times 440 \times h$[a] | 407 × 407 × 629 |
| Voxel size (mm$^3$/voxel) | 2.04×2.04×2.03 | 2.04×2.04×2.03 | 1.65×1.65×1.65 | 1.67×1.67×2.89 |
| Acquisition details | Single pass of CBM acquisition, with 5 mins/bed position after 60-min post-injection tracer uptake time. | CBM 19-pass scanning over 90 mins, starting immediately after tracer injection. High-count images were reconstructed by combining list-mode data of all passes acquired after 60-min post-injection (Liu et al., 2022). | Similar setting with Dataset A-Group II. Standard-count images were reconstructed from a 5-min single pass. | 20-min scanning after 120-min post-injection tracer uptake time. |
| Count levels | 5,10,20%, standard-count | 5,40%, high-count | standard-, high-count | 2.5,6.25,12.5%, high-count |
| # subjects | 195 | 29 | 10 | 10 |
| Lesion sizes (mean ± std,mL) | 7.95 ± 74.23 (min:0.025, max:2542.68) | 12.95 ± 39.74 (min:0.017, max:245.62) | 6.34 ± 17.28 (min:0.009, max:87.44) | 0.69 ± 0.78 (min:0.04, max:2.83) |
| Lesion locations | Tumors in the liver, lungs, bowel, bones, peritoneum, neck, para-rectal soft tissue, abdomen, thyroid, breast, pancreas, skin, colon, and rectum. Lymph nodes in the para-aortic, inguinal, mediastinal, diaphragmatic, hilar, and cervical regions. | Tumors in the pulmonary hilum, lung, and pleura; Lymph nodes in the pectoral, axillary, mediastinal, thoracic, para-aortic, inguinal, and external iliac regions. | Tumors in the lungs, neck, and bowel; Lymph nodes in the para-aortic region. | Tumors in the neck, lungs, and bones |
| Train/val/test | 175/20/0 | 0/0/29 | 0/0/10 | 3/2/5 |

[a] $h$ depends on patient height.

**Table 2**
Denoising comparisons with SOTA methods. Metrics are averaged across all noise levels and presented as mean ± std. A lower value is preferred, with the best result in each case highlighted in **bold**.

| Scanner | Method | NRMSE on whole image and each semantic class | | | | | | | | |
|---|---|---|---|---|---|---|---|---|---|---|
| | | Whole image | Lesion | Liver | Lung | Bone | Muscle | Kidney | Spleen | Aorta |
| mCT | LeqMod-GAN | .156 ± .144 | .151 ± .106 | .125 ± .059 | .156 ± .067 | .160 ± .085 | .169 ± .081 | .169 ± .109 | .132 ± .064 | .134 ± .062 |
| | SpachTransformer | .149 ± .160 | .160 ± .134 | .086 ± .031 | .123 ± .044 | .132 ± .079 | **.142 ± .070** | .155 ± .112 | **.102 ± .049** | .111 ± .047 |
| | 3D Med-DDPM | .153 ± .123 | .232 ± .306 | .105 ± .040 | .159 ± .061 | .132 ± .149 | .147 ± .107 | .152 ± .073 | .119 ± .058 | .132 ± .050 |
| | AMDiff | **.146 ± .116**[*] | **.150 ± .172** | **.080 ± .041**[*] | **.115 ± .053**[*] | **.129 ± .131**[*] | .148 ± .106 | **.148 ± .121**[*] | .109 ± .056 | **.101 ± .049**[*] |
| Vision | LeqMod-GAN | .193 ± .056 | .134 ± .055 | .103 ± .012 | .102 ± .021 | .136 ± .046 | .123 ± .034 | .156 ± .047 | .091 ± .028 | .122 ± .011 |
| | SpachTransformer | .192 ± .056 | .136 ± .052 | .080 ± .012 | .093 ± .022 | .139 ± .048 | .117 ± .035 | **.145 ± .050** | **.072 ± .027** | .100 ± .013 |
| | 3D Med-DDPM | .189 ± .060 | .201 ± .084 | .086 ± .010 | .138 ± .027 | .142 ± .046 | .120 ± .034 | .175 ± .061 | .088 ± .028 | .110 ± .018 |
| | AMDiff | **.183 ± .058**[*] | **.132 ± .052** | **.080 ± .011** | **.090 ± .023** | **.133 ± .046**[*] | **.116 ± .033**[*] | .147 ± .053 | .075 ± .028 | **.097 ± .014**[*] |
| uExplorer | LeqMod-GAN | .155 ± .065 | .265 ± .096 | .169 ± .040 | .217 ± .048 | .202 ± .033 | .208 ± .050 | .180 ± .033 | .161 ± .041 | .203 ± .047 |
| | SpachTransformer | .134 ± .033 | .287 ± .125 | .139 ± .035 | **.184 ± .041** | .176 ± .041 | .181 ± .047 | **.175 ± .044** | .141 ± .041 | .188 ± .050 |
| | 3D Med-DDPM | .133 ± .024 | .266 ± .144 | .134 ± .015 | .239 ± .049 | .182 ± .023 | .180 ± .032 | .207 ± .025 | .139 ± .024 | .195 ± .021 |
| | AMDiff | **.130 ± .026**[*] | **.250 ± .123**[*] | **.126 ± .016**[*] | .188 ± .026 | **.172 ± .025** | **.177 ± .027**[*] | .182 ± .026 | **.125 ± .024**[*] | **.180 ± .022**[*] |

[*] P-value < 0.05 based on the non-parametric Wilcoxon signed-rank test between the AMDiff and others.

Results are summarized in Table 3. SwinUNETR, nnMamba, and MedSAM achieved satisfactory Dice coefficients for major organs, with average scores of 0.85/0.87, 0.84/0.88, and 0.85/0.88 for the liver/lung, respectively, averaged across all test cases and scanners. AMDiff, benefiting from its integrated denoising component and joint training framework, achieved the highest overall performance, with average Dice coefficients of 0.86/0.89 for the liver/lung. For lesion segmentation, SwinUNETR, nnMamba, and MedSAM achieved average Dice scores of 0.47, 0.47, and 0.48, respectively. AMDiff yielded a modest improvement, achieving an average Dice coefficient of 0.52. The relatively low Dice scores across all methods are primarily due to the inherent difficulty of detecting lesions given only low-count PET images, as well as the predominance of small lesions in our test set (see Table 1), which adversely impacts Dice scores.

Fig. 5 and 6 present visual comparisons of lesion and organ segmentation, respectively. Segmentation masks obtained by SwinUNETR, nnMamba, and Med-SAM sometimes fail to capture small objects, such as lesions and ribs. In contrast, the AMDiff, with its denoiser enhancing object visibility, demonstrates superior performance in distinguishing lesions and anatomical structures.

### 4.3. Ablation studies

To evaluate the individual contributions of key components in AMDiff and corroborate the synergies between denoising and segmentation, ablation studies were conducted on the denoising revision module, lesion regularizer, organ regularizer, and the ResMamba module. The NRMSE and Dice metrics for AMDiff and its degraded variants, averaged across all test cases, are presented in Table 4.

The revision module, serving as a bridge for transferring denoised information to the segmenter, substantially improved lesion segmentation accuracy, raising the Dice coefficient for the lesion class by

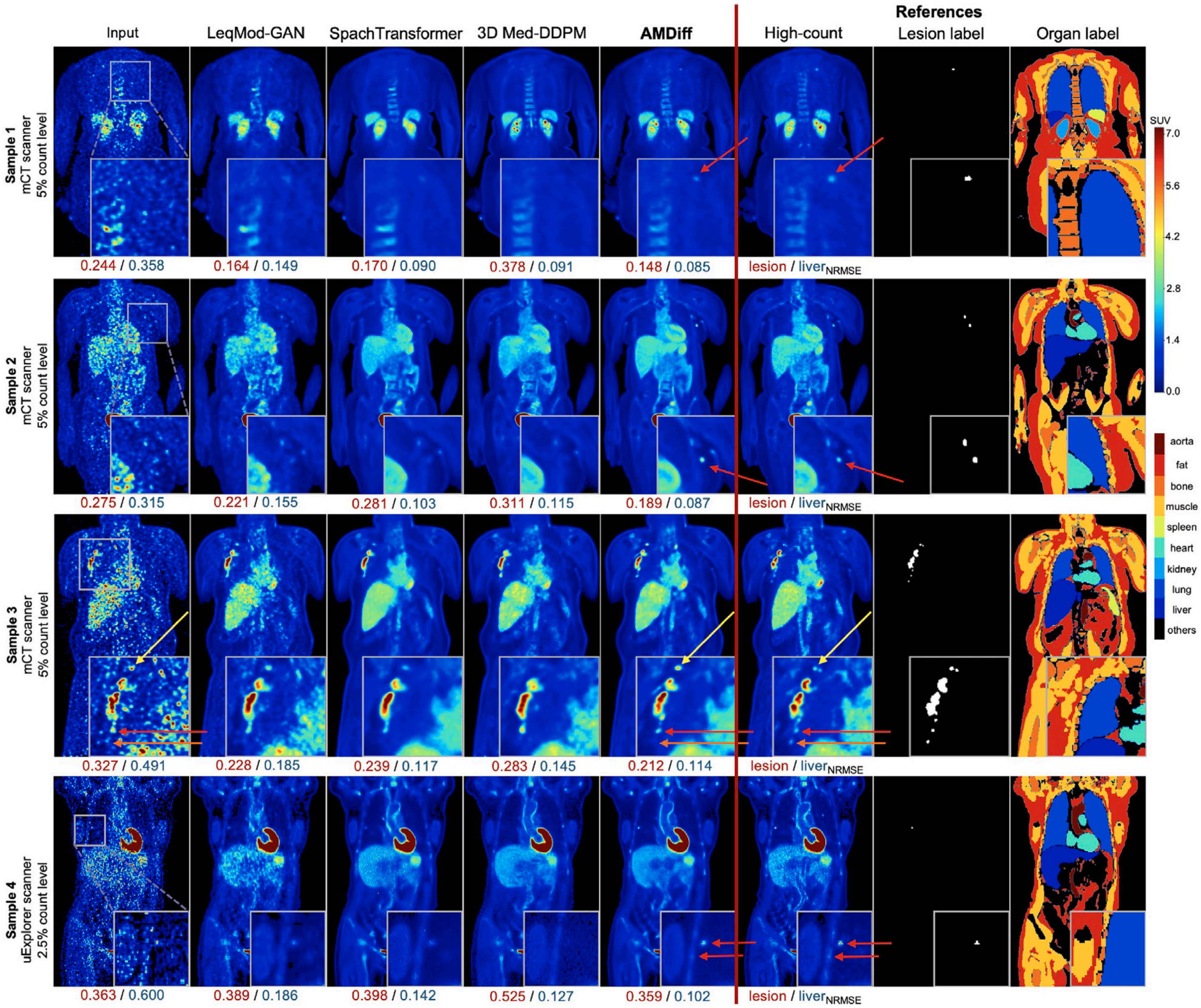

**Fig. 4.** Visual comparisons between AMDiff and other denoising models. NRMSE metrics for the lesion and liver are displayed below each image. ROIs are cropped, magnified, and shown in the bottom-right corner of each image. Red arrows highlight areas where lesions and organ structures in AMDiff-denoised results appear superior to those in comparison models and more closely resemble high-count references. Yellow arrows indicate lesions in AMDiff-denoised images that exhibit slight differences from high-count references but still outperform comparison models and remain consistent with their appearance in the input. Orange arrows denote small, weak lesions that are missed in all denoised outputs, likely due to their extremely faint signals and the challenge of distinguishing them from high noise in low-count inputs.

**Table 3**
Segmentation comparisons with SOTA methods. Inputs to the comparison segmentation models are concatenation of low-count images and denoised ones obtained by the conditioned diffusion module architecture used in AMDiff. Metrics are averaged across all noise levels and presented as mean ± std. Higher Dice scores are preferred, with the best result in each case highlighted in bold.

| Scanner | Method | Dice on each semantic class | | | | | | | |
|---|---|---|---|---|---|---|---|---|---|
| | | Lesion | Liver | Lung | Bone | Muscle | Kidney | Spleen | Aorta |
| mCT | SwinUNETR | .509 ± .461 | .854 ± .057 | .893 ± .038 | .584 ± .057 | .764 ± .046 | .553 ± .160 | .573 ± .232 | .719 ± .090 |
| | nnMamba | .526 ± .330 | .850 ± .061 | .905 ± .029 | .618 ± .066 | .774 ± .047 | .554 ± .164 | .575 ± .226 | .738 ± .078 |
| | Med-SAM | .531 ± .336 | **.865 ± .049** | .903 ± .037 | .628 ± .059 | .773 ± .044 | .591 ± .155 | .570 ± .230 | .630 ± .109 |
| | AMDiff | **.573 ± .251*** | .862 ± .055 | **.912 ± .033*** | **.651 ± .063*** | **.780 ± .047*** | **.596 ± .166** | **.606 ± .270** | **.757 ± .079*** |
| Vision | SwinUNETR | .449 ± .463 | .709 ± .049 | .768 ± .148 | .519 ± .066 | .742 ± .041 | .742 ± .098 | .492 ± .160 | .422 ± .132 |
| | nnMamba | .438 ± .231 | .690 ± .062 | .810 ± .093 | .549 ± .064 | .742 ± .042 | .719 ± .145 | .506 ± .081 | .434 ± .134 |
| | Med-SAM | .440 ± .323 | .715 ± .056 | .815 ± .101 | .554 ± .075 | **.764 ± .044** | .726 ± .121 | .500 ± .131 | .450 ± .158 |
| | AMDiff | **.466 ± .290*** | **.752 ± .069*** | **.819 ± .103** | **.596 ± .090*** | .747 ± .034 | **.772 ± .086*** | **.673 ± .166*** | **.610 ± .133*** |
| uExplorer | SwinUNETR | .386 ± .363 | .894 ± .016 | .847 ± .080 | .483 ± .063 | .728 ± .037 | .748 ± .054 | .787 ± .050 | .439 ± .135 |
| | nnMamba | .338 ± .336 | .888 ± .112 | .839 ± .095 | .520 ± .108 | .693 ± .047 | .759 ± .067 | .781 ± .074 | .591 ± .089 |
| | Med-SAM | .368 ± .318 | .899 ± .021 | .859 ± .062 | .563 ± .066 | .732 ± .043 | .731 ± .079 | **.825 ± .046** | .603 ± .127 |
| | AMDiff | **.396 ± .362*** | **.901 ± .034** | **.879 ± .049*** | **.633 ± .062*** | **.746 ± .045** | **.815 ± .040*** | .794 ± .129 | **.758 ± .071*** |

* P-value < 0.05 based on the non-parametric Wilcoxon signed-rank test between the AMDiff and others.

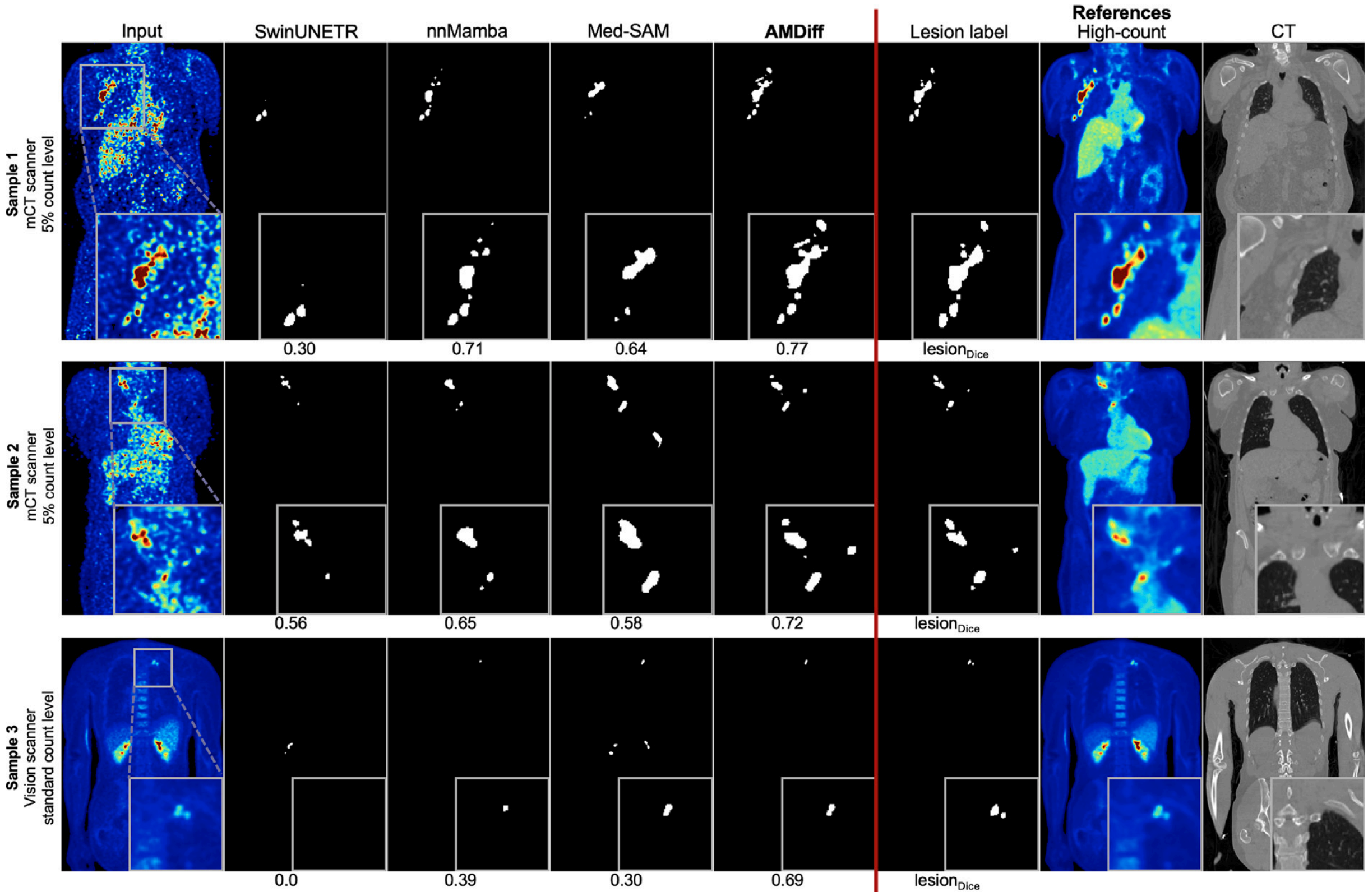

**Fig. 5.** Visual comparisons of AMDiff with other models for lesion segmentation. Dice coefficients are shown below each image. ROIs are cropped, magnified, and displayed in the bottom-right corner of each image.

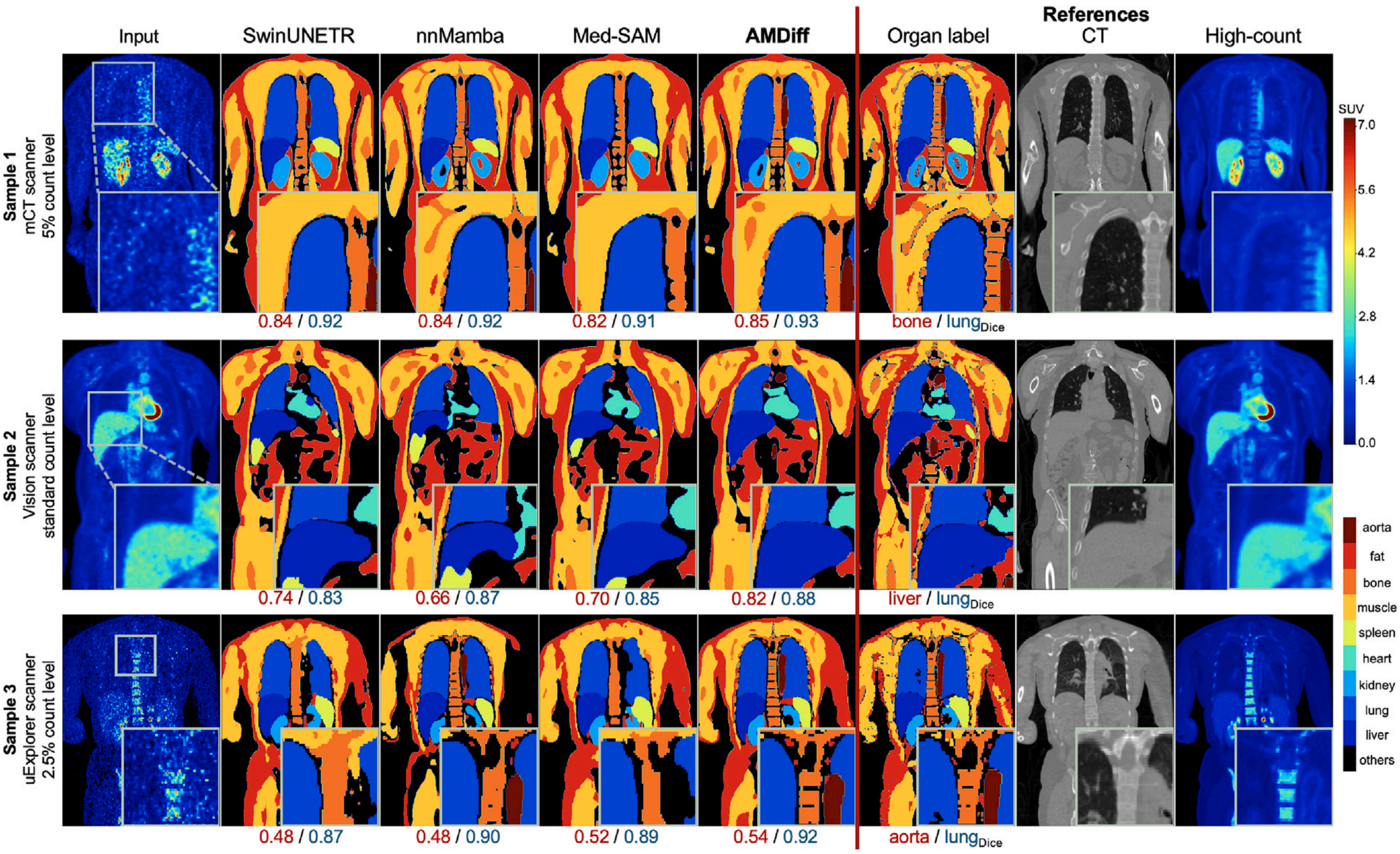

**Fig. 6.** Visual comparisons of AMDiff with other models for organ segmentation. Dice coefficients for organs are shown below each image. ROIs are cropped, magnified, and displayed in the bottom-right corner of each image.

**Table 4**
Ablation studies on the AMDiff. Metrics are averaged across all test cases and are presented as mean ± std. Lower NRMSE values and higher Dice scores indicate better performance. The best result for each case is highlighted in bold.

| Method | NRMSE on whole image and each semantic class | | | | | | | | | Dice on each semantic class | | | | | | | |
|---|---|---|---|---|---|---|---|---|---|---|---|---|---|---|---|---|---|
| | Whole | Lesion | Liver | Lung | Bone | Muscle | Kidney | Spleen | Aorta | Lesion | Liver | Lung | Bone | Muscle | Kidney | Spleen | Aorta |
| AMDiff | **.147 ±.097** | **.185 ±.149** | **.091 ±.039** | **.130 ±.056** | .139 ± .109 | **.152 ±.089** | **.156 ±.101** | **.110 ±.050** | .119 ± .054 | **.520 ±.216** | **.860 ±.066** | **.894 ±.056** | **.641 ±.047** | **.768 ±.048** | **.666 ±.167** | **.658 ±.253** | **.742 ±.133** |
| w/o denoising revision | .163 ± .099 | .217 ± .171 | .093 ± .040 | .137 ± .058 | .155 ± .120 | .167 ± .097 | .174 ± .113 | .113 ± .054 | .123 ± .055 | .409 ± .214 | .837 ± .062 | .881 ± .053 | .591 ± .066 | .752 ± .046 | .631 ± .173 | .626 ± .243 | .702 ± .089 |
| w/o organ regularizer | .156 ± .105 | .186 ± .172 | .107 ± .038 | .134 ± .070 | .156 ± .115 | .166 ± .098 | .163 ± .118 | .114 ± .057 | .128 ± .052 | .519 ± .225 | .855 ± .074 | .882 ± .061 | .632 ± .064 | .763 ± .045 | .652 ± .169 | .652 ± .251 | .731 ± .120 |
| w/o lesion regularizer | .147 ± .103 | .274 ± .176 | .095 ± .037 | .130 ± .058 | **.137 ±.116** | .153 ± .100 | .158 ± .109 | .111 ± .053 | **.116 ±.051** | .474 ± .218 | .860 ± .068 | .892 ± .057 | .640 ± .075 | .766 ± .045 | .662 ± .176 | .655 ± .248 | .740 ± .128 |
| w/o ResMamba | .149 ± .115 | .189 ± .161 | .098 ± .034 | .130 ± .057 | .143 ± .127 | .158 ± .102 | .161 ± .113 | .112 ± .050 | .121 ± .043 | .476 ± .206 | .849 ± .069 | .884 ± .054 | .611 ± .069 | .761 ± .045 | .646 ± .172 | .638 ± .247 | .701 ± .129 |

27.1%. It also benefited organ segmentation, albeit to a lesser degree, as organs are generally more distinguishable than lesions in noisy images. Specifically, the module improved Dice scores for the liver, lung, bone, and aorta by 2.7%, 1.5%, 8.5%, and 5.7%, respectively. In addition, the revision module contributed to improved denoising performance, benefiting from the coupled effect between denoising and segmentation, whereby enhanced segmentation demands the generation of more structurally accurate denoised outputs.

The lesion-organ-specific regularizer, which employs semantic labels to supervise the denoising process, proved effective in constraining the appearance of lesions and organs in the denoised outputs. The lesion regularizer reduced lesion NRMSE by 32.5% and, through the cascading effect of improved denoising on downstream tasks, increased the lesion Dice coefficient by 9.7%. Similarly, the organ regularizer decreased NRMSE for bone and aorta by 10.9% and 7.0%, respectively, while also slightly improving the corresponding Dice coefficients by 1.4% and 1.5%.

Additionally, we performed an ablation study by replacing the ResMamba module with standard convolutional layers, thereby degrading the segmenter architecture to a nnU-Net. As shown in Table 4, incorporating the ResMamba module led to improved segmentation performance across all semantic classes. This enhancement is attributed to ResMamba's superior ability to capture both local and long-range contextual dependencies, as well as its built-in feature augmentation mechanisms. Notably, Dice scores for the lesion, bone, and aorta classes increased by 9.2%, 4.9%, and 5.8%, respectively. In addition, the inclusion of ResMamba also yielded modest gains in denoising performance, likely due to the coupled effect between segmentation and denoising.

### 4.4. Clinical metric quantification

With simultaneous denoising and segmentation, the AMDiff enables one-step automatic quantification of clinical metrics directly from low-count inputs. To evaluate the joint model as a whole in quantification accuracy, the MTV, TLG, and organ $SUV_{mean}$ values computed by AMDiff are compared against ground truth values derived from high-count images and manual annotations using linear regression analysis, with results shown in Fig. 7.

The AMDiff demonstrates high correlations with $R^2$ values of 0.98/0.98 for liver/aorta $SUV_{mean}$, and promising results of 0.81/0.88 for MTV/TLG quantification. Both the lesion-organ-specific regularizer and the denoising revision module contribute to improved agreement between automatic quantification and ground truth, as illustrated in Fig. 7.

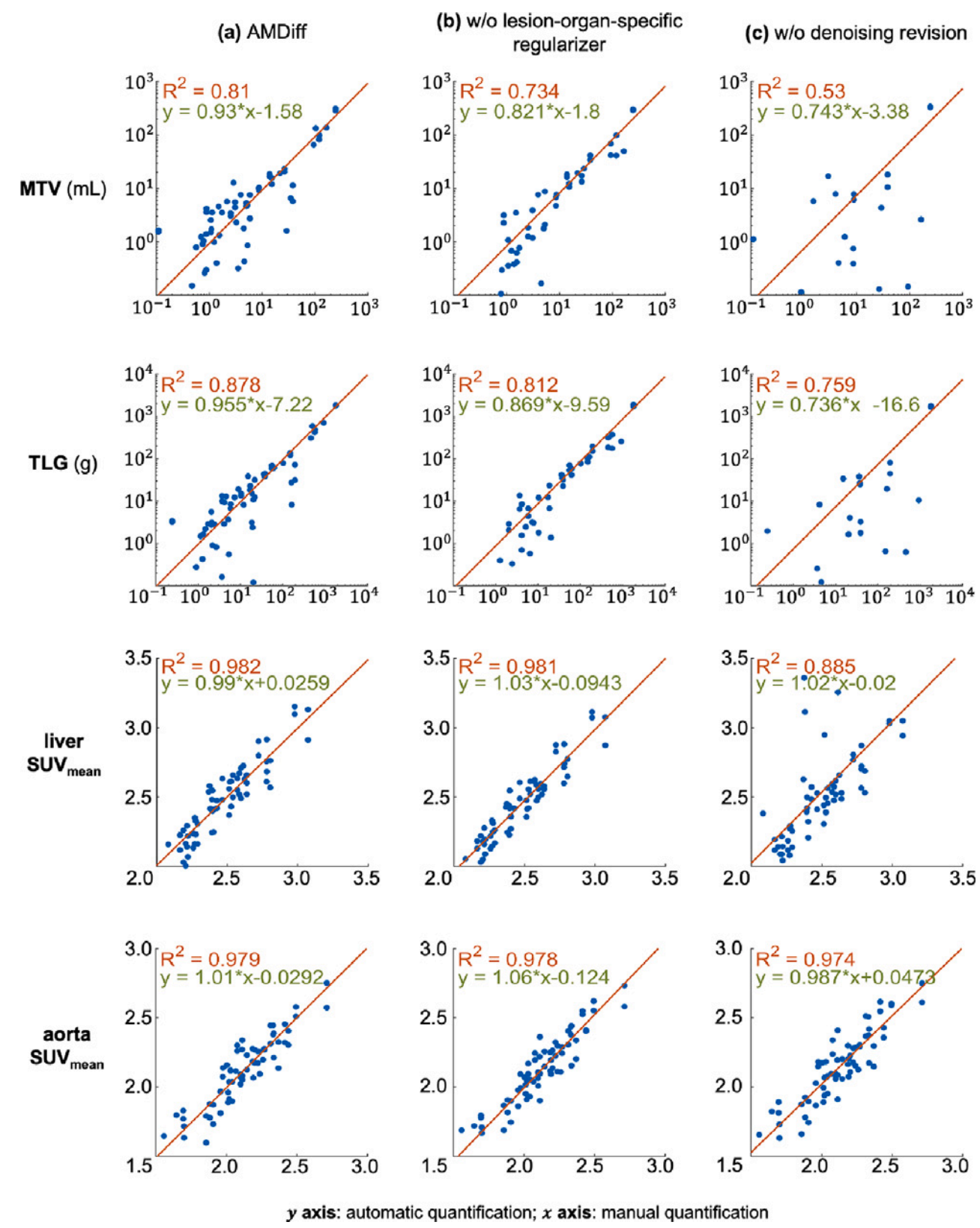

**Fig. 7.** Linear regression analysis on clinical metric quantification, including MTV, TLG, and organ $SUV_{mean}$, taking the liver and aorta as examples.

## 5. Discussion

The results demonstrate the feasibility and advantage of simultaneously performing denoising and lesion/organ segmentation in low-count PET imaging, with denoising aimed at potential reductions in radiotracer dose and scan time, and auto-segmentation intended to alleviate physician workload. By effectively leveraging synergies between these tasks, the proposed AMDiff model enables automatic quantification of clinical metrics directly from low-count noisy PET images, achieving good agreement with ground truth values. In contrast, existing PET image analysis methods, either for low-count image denoising or standard-count image segmentation, cannot independently quantify metrics such as TLG from low-count inputs.

The AMDiff model, comprising a semantic-informed denoiser and a denoising-informed segmenter, is distinguished by its inherent bidirectional information exchange between the denoiser and segmenter, as demonstrated in Table 4 and Fig. 7. In contrast, similar approaches exploring inter-task synergies (Xia et al., 2024; Ye and Xu, 2024; Peng et al., 2023) often employ unidirectional strategies, where one task aids the other, either segmentation supporting denoising or vice versa. These methods typically depend on an independent pre-trained model to provide priors for the main task, introducing potential biases if the priors are inaccurate. AMDiff addresses these limitations by integrating denoising and segmentation into a unified framework. Although its internal structure follows a sequential denoiser-then-segmenter architecture, the inclusion of the denoising revision module establishes an effective link between the two tasks. Combined with joint optimization, this design facilitates bidirectional interaction between the denoiser and segmenter. Furthermore, both components have direct access to original low-count inputs and are supervised by ground-truth labels. This eliminates the need for separate prior models, thereby reducing potential biases and simplifying the overall system.

The denoising revision module facilitates the flow of information from the denoiser to the segmenter, resulting in more accurate segmentation masks for both lesions and organs. As shown in Table 4, this module improved the Dice coefficients for lesions and aorta by 27.14% and 5.70%, respectively, averaged across all test cases. Additionally, it contributed to improved denoising performance due to the cascading and bidirectional interactions between tasks, where better segmentation relies on higher-quality denoised inputs. Specifically, lesion and aorta NRMSE values were reduced by 14.75% and 3.25%, respectively. As a final result, these improvements in both segmentation and denoising led to more accurate quantification of organ $SUV_{mean}$ and the TLG, as validated in Fig. 7. By introducing this intermediate connection step, AMDiff outperformed other segmentation models, including SwinUNETR (He et al., 2023), nnMamba (Gong et al., 2024b), and Med-SAM (Wang et al., 2024), across nearly all semantic classes, as summarized in Table 3. It exhibited greater robustness to heavy input noise and demonstrates superior performance in segmenting small lesions and bone structures, as illustrated in Fig. 5 and 6.

The lesion-organ-specific regularizer, which incorporates additional segmentation labels to supervise the denoised outputs, enhanced both lesion-wise and organ-wise SUV consistency between the denoised results and high-count references, as shown in Table 4. This contributed to improved anatomical and metabolic reliability, reducing NRMSE for lesions and lungs by 32.48% and 2.99%, respectively, averaged across all test cases. Furthermore, due to connections between the denoiser and segmenter where improvements in one task trigger positive cascading effects on the other, the regularizer also contributed to increased Dice coefficients. As a result, the overall improvement in denoising and segmentation led to more accurate quantification of clinical metrics such as TLG, as validated in Fig. 7. Additionally, the use of semantic regularization enabled AMDiff to outperform representative denoising methods, including LeqModGAN (Xia et al., 2024), SpachTranformer (Jang et al., 2023), and Med-DDPM (Dorjsembe et al., 2024), particularly in recovering small lesions and bones, as demonstrated in Fig. 4.

While the multi-task AMDiff model demonstrates promising performance, there remains room for further improvement. In lesion segmentation, all lesion types were treated as a single class due to the limited size and diversity of available datasets. Future work could benefit from expanding the dataset to include a broader range of lesion locations, sizes, and cancer types and stages. Exploring lesion-type-specific segmentation may enhance quantification accuracy by incorporating more detailed lesion characteristics. For organ segmentation, organ labels were generated using the TotalSegmentator tool, with manual corrections applied to address obvious errors, ensuring general accuracy. However, occasional label inaccuracies may still persist, potentially affecting both model performance and evaluation outcomes. Future research could incorporate higher-quality organ annotations to further enhance segmentation accuracy and the robustness of experimental results. Regarding denoising, a relatively small number of diffusion timesteps (250) was used to maintain acceptable inference time. However, processing a single image of $440 \times 440 \times 620$ voxels still required approximately 26.6 min for denoising and segmentation. While increasing the number of timesteps could potentially enhance performance, it would further increase the computational burden. Future work should explore the use of more efficient strategies, such as latent diffusion models (Kim and Park, 2024), to strike a better balance between performance and practicality.

## 6. Conclusion

We present AMDiff, a unified model for denoising and lesion/organ segmentation in low-count PET imaging. By simultaneously mapping low-count images to high-count equivalents and generating lesion and organ masks, AMDiff facilitates one-step clinical metric quantification, offering practical advantages. The model effectively leverages synergies between denoising and segmentation tasks, as validated through extensive experiments. With the dual tasks mutually enhancing each other, AMDiff outperforms SOTA denoising and segmentation benchmarks.

## CRediT authorship contribution statement

**Menghua Xia:** Writing – original draft, Methodology, Formal analysis, Data curation, Conceptualization. **Kuan-Yin Ko:** Data curation. **Der-Shiun Wang:** Data curation. **Ming-Kai Chen:** Data curation. **Qiong Liu:** Resources, Investigation. **Huidong Xie:** Resources, Investigation. **Liang Guo:** Resources, Investigation. **Wei Ji:** Visualization, Validation. **Jinsong Ouyang:** Writing – review & editing, Validation. **Reimund Bayerlein:** Writing – review & editing, Data curation. **Benjamin A. Spencer:** Writing – review & editing, Data curation. **Quanzheng Li:** Writing – review & editing, Validation. **Ramsey D. Badawi:** Writing – review & editing, Validation. **Georges El Fakhri:** Writing – review & editing, Supervision, Project administration. **Chi Liu:** Writing – review & editing, Project administration, Funding acquisition, Conceptualization.

## Declaration of competing interest

The authors declare that they have no known competing financial interests or personal relationships that could have appeared to influence the work reported in this paper.

## Acknowledgments

This work was supported by the National Institutes of Health, United States (NIH) under grants R01CA275188 and R01EB025468, and the National Institute of Biomedical Imaging and Bioengineering (NIBIB) under grant P41EB022544.

## References

Ahamed, S., Tek, M., Kurkowska, S., Uribe, C., Rahmim, A., 2024. Leveraging counterfactual generative diffusion probabilistic model for anomaly detection: Application to lung cancer PET images. J. Nucl. Med. 65 (supplement 2), 241881–241881.

Angelis, G., Fuller, O., Gillam, J., Meikle, S., 2021. Denoising non-steady state dynamic PET data using a feed-forward neural network. Phys. Med. Biol. 66 (3), 034001.

Bousse, A., Kandarpa, V.S.S., Shi, K., Gong, K., Lee, J.S., Liu, C., Visvikis, D., 2024. A review on low-dose emission tomography post-reconstruction denoising with neural network approaches. IEEE Trans. Radiat. Plasma Med. Sci..

Chen, H.H., Chiu, N.-T., Su, W.-C., Guo, H.-R., Lee, B.-F., 2012. Prognostic value of whole-body total lesion glycolysis at pretreatment FDG PET/CT in non–small cell lung cancer. Radiology 264 (2), 559–566.

Cui, J., Gong, K., Guo, N., Wu, C., Meng, X., Kim, K., Zheng, K., Wu, Z., Fu, L., Xu, B., Zhu, Z., Tian, J., Liu, H., Li, Q., 2019. PET image denoising using unsupervised deep learning. Eur. J. Nucl. Med. Mol. Imaging 46, 2780–2789.

Dorjsembe, Z., Pao, H.-K., Odonchimed, S., Xiao, F., 2024. Conditional diffusion models for semantic 3D brain MRI synthesis. IEEE J. Biomed. Heal. Inform. 28 (7), 4084–4093.

Fu, M., Zhang, N., Huang, Z., Zhou, C., Zhang, X., Yuan, J., He, Q., Yang, Y., Zheng, H., Liang, D., Wu, F.-X., Fan, W., Hu, Z., 2024. OIF-Net: An optical flow registration-based PET/MR cross-modal interactive fusion network for low-count brain PET image denoising. IEEE Trans. Med. Imaging 43 (4), 1554–1567.

Gatidis, S., Früh, M., Fabritius, M.P., Gu, S., Nikolaou, K., Fougère, C.L., Ye, J., He, J., Peng, Y., Bi, L., Ma, J., Wang, B., Zhang, J., Huang, Y., Heiliger, L., Marinov, Z., Stiefelhagen, R., Egger, J., Kleesiek, J., Sibille, L., Xiang, L., Bendazzoli, S., Astaraki, M., Ingrisch, M., Cyran, C.C., Küstner, T., 2024. Results from the autopet challenge on fully automated lesion segmentation in oncologic PET/CT imaging. Nat. Mach. Intell. 1–10.

Gatidis, S., Hepp, T., Früh, M., La Fougère, C., Nikolaou, K., Pfannenberg, C., Schölkopf, B., Küstner, T., Cyran, C., Rubin, D., 2022. A whole-body FDG-PET/CT dataset with manually annotated tumor lesions. Sci. Data 9 (1), 601.

Gong, K., Johnson, K., El Fakhri, G., Li, Q., Pan, T., 2024a. PET image denoising based on denoising diffusion probabilistic model. Eur. J. Nucl. Med. Mol. Imaging 51 (2), 358–368.

Gong, H., Kang, L., Wang, Y., Wan, X., Li, H., 2024b. Nnmamba: 3D biomedical image segmentation, classification and landmark detection with state space model. arXiv preprint arXiv:2402.03526.

Han, Z., Wang, Y., Zhou, L., Wang, P., Yan, B., Zhou, J., Wang, Y., Shen, D., 2023. Contrastive diffusion model with auxiliary guidance for coarse-to-fine PET reconstruction. In: International Conference on Medical Image Computing and Computer-Assisted Intervention. Springer, pp. 239–249.

He, Y., Nath, V., Yang, D., Tang, Y., Myronenko, A., Xu, D., 2023. Swinunetr-v2: stronger swin transformers with stagewise convolutions for 3D medical image segmentation. In: International Conference on Medical Image Computing and Computer-Assisted Intervention. Springer, pp. 416–426.

Ho, J., Jain, A., Abbeel, P., 2020. Denoising diffusion probabilistic models. Adv. Neural Inf. Process. Syst. 33, 6840–6851.

Hu, R., Yoon, S., Wu, D., Tivnan, M., Chen, Z., Wang, Y., Luo, J., Cui, J., Li, Q., Liu, H., Guo, N., 2024. Realistic tumor generation using 3D conditional latent diffusion model. J. Nucl. Med. 65 (supplement 2), 241725–241725.

Huang, Z., Liu, Z., He, P., Ren, Y., Li, S., Lei, Y., Luo, D., Liang, D., Shao, D., Hu, Z., et al., 2022. Segmentation-guided denoising network for low-dose CT imaging. Comput. Methods Programs Biomed. 227, 107199.

Jang, S.-I., Pan, T., Li, Y., Heidari, P., Chen, J., Li, Q., Gong, K., 2023. Spach-Transformer: spatial and channel-wise transformer based on local and global self-attentions for PET image denoising. IEEE Trans. Med. Imaging.

Kim, J., Park, H., 2024. Adaptive latent diffusion model for 3D medical image to image translation: Multi-modal magnetic resonance imaging study. In: Proceedings of the IEEE/CVF Winter Conference on Applications of Computer Vision. pp. 7604–7613.

Kitson, S.L., Cuccurullo, V., Ciarmiello, A., Salvo, D., Mansi, L., 2009. Clinical applications of positron emission tomography (PET) imaging in medicine: oncology, brain diseases and cardiology. Curr. Radiopharm. 2 (4), 224–253.

Leung, K.H., Rowe, S.P., Sadaghiani, M.S., Leal, J.P., Mena, E., Choyke, P.L., Du, Y., Pomper, M.G., 2024. Deep semisupervised transfer learning for fully automated whole-body tumor quantification and prognosis of cancer on PET/CT. J. Nucl. Med. 65 (4), 643–650.

Li, X., Ding, M., Gu, Y., Pižurica, A., 2023. An end-to-end framework for joint denoising and classification of hyperspectral images. IEEE Trans. Neural Netw. Learn. Syst. 34 (7), 3269–3283.

Liebgott, A., Lorenz, C., Gatidis, S., Vu, V.C., Nikolaou, K., Yang, B., 2021. Automated multi-organ segmentation in PET images using cascaded training of a 3D U-net and convolutional autoencoder. In: IEEE International Conference on Acoustics, Speech and Signal Processing. IEEE, pp. 1145–1149.

Lim, H., 2023. Transformer-based integrated framework for joint reconstruction and segmentation in accelerated knee MRI. Electronics 12 (21), 4434.

Liu, J., Ren, S., Wang, R., Mirian, N., Tsai, Y.-J., Kulon, M., Pucar, D., Chen, M.-K., Liu, C., 2022. Virtual high-count PET image generation using a deep learning method. Med. Phys. 49 (9), 5830–5840.

Onishi, Y., Hashimoto, F., Ote, K., Ohba, H., Ota, R., Yoshikawa, E., Ouchi, Y., 2021. Anatomical-guided attention enhances unsupervised PET image denoising performance. Med. Image Anal. 74, 102226.

Peng, D., Hu, P., Ke, Q., Liu, J., 2023. Diffusion-based image translation with label guidance for domain adaptive semantic segmentation. In: Proceedings of the IEEE/CVF International Conference on Computer Vision. pp. 808–820.

Salimi, Y., Mansouri, Z., Shiri, I., Mainta, I., Zaidi, H., 2022. Deep learning-powered CT-less multitracer organ segmentation from pet images: a solution for unreliable CT segmentation in PET/CT imaging. Clin. Nucl. Med. 10–1097.

Shiyam Sundar, L.K., Yu, J., Muzik, O., Kulterer, O.C., Fueger, B., Kifjak, D., Nakuz, T., Shin, H.M., Sima, A.K., Kitzmantl, D., Badawi, R.D., Nardo, L., Cherry, S.R., Spencer, B.A., Hacker, M., Beyer, T., 2022. Fully automated, semantic segmentation of whole-body 18F-FDGPET/CT images based on data-centric artificial intelligence. J. Nucl. Med. 63 (12), 1941–1948.

Suganuma, Y., Teramoto, A., Saito, K., Fujita, H., Suzuki, Y., Tomiyama, N., Kido, S., 2023. Hybrid multiple-organ segmentation method using multiple U-Nets in PET/CT images. Appl. Sci. 13 (19), 10765.

Wang, H., Guo, S., Ye, J., Deng, Z., Cheng, J., Li, T., Chen, J., Su, Y., Huang, Z., Shen, Y., Fu, B., Zhang, S., He, J., Qiao, Y., 2024. SAM-Med3D: towards general-purpose segmentation models for volumetric medical images. Preprint at https://arxiv.org/abs/2310.15161.

Wasserthal, J., Breit, H.-C., Meyer, M.T., Pradella, M., Hinck, D., Sauter, A.W., Heye, T., Boll, D.T., Cyriac, J., Yang, S., Bach, M., Segeroth, M., 2023. TotalSegmentator: robust segmentation of 104 anatomic structures in CT images. Radiol.: Artif. Intell. 5 (5).

Xia, M., Bayerlein, R., Chemli, Y., Liu, X., Ouyang, J., Fakhri, G.E., Badawi, R.D., Li, Q., Liu, C., 2025. DREAM: On hallucinations in AI-generated content for nuclear medicine imaging. arXiv preprint arXiv:2506.13995.

Xia, M., Xie, H., Liu, Q., Zhou, B., Wang, H., Li, B., Rominger, A., Shi, K., Fakhri, G.E., Liu, C., 2024. Lpqcm: adaptable lesion-quantification-consistent modulation for deep learning low-count PET image denoising. arXiv preprint arXiv:2404.17994.

Xie, H., Gan, W., Zhou, B., Chen, M.-K., Kulon, M., Boustani, A., Spencer, B.A., Bayerlein, R., Ji, W., Chen, X., Liu, Q., Guo, X., Xia, M., Zhou, Y., Liu, H., Guo, L., An, H., Kamilov, U.S., Wang, H., Li, B., Rominger, A., Shi, K., Wang, G., Badawi, R.D., Liu, C., 2024. Dose-aware diffusion model for 3D low-dose PET: multi-institutional validation with reader study and real low-dose data. arXiv preprint arXiv:2405.12996.

Xie, H., Liu, Q., Zhou, B., Chen, X., Guo, X., Wang, H., Li, B., Rominger, A., Shi, K., Liu, C., 2023. Unified noise-aware network for low-count PET denoising with varying count levels. IEEE Trans. Radiat. Plasma Med. Sci..

Xu, S., Sun, K., Liu, D., Xiong, Z., Zha, Z.-J., 2023. Synergy between semantic segmentation and image denoising via alternate boosting. ACM Trans. Multimed. Comput. Commun. Appl. 19 (2), 1–23.

Xue, S., Guo, R., Bohn, K.P., Matzke, J., Viscione, M., Alberts, I., Meng, H., Sun, C., Zhang, M., Zhang, M., Sznitman, R., El Fakhri, G., Rominger, A., Li, B., Shi, K., 2022. A cross-scanner and cross-tracer deep learning method for the recovery of standard-dose imaging quality from low-dose PET. Eur. J. Nucl. Med. Mol. Imaging 1–14.

Ye, H., Xu, D., 2024. Diffusionmtl: learning multi-task denoising diffusion model from partially annotated data. In: Proceedings of the IEEE/CVF Conference on Computer Vision and Pattern Recognition. pp. 27960–27969.

Yeung, M., Sala, E., Schönlieb, C.-B., Rundo, L., 2022. Unified focal loss: generalising dice and cross entropy-based losses to handle class imbalanced medical image segmentation. Comput. Med. Imaging Graph. 95, 102026.

Yousefirizi, F., Jha, A.K., Brosch-Lenz, J., Saboury, B., Rahmim, A., 2021. Toward high-throughput artificial intelligence-based segmentation in oncological PET imaging. PET Clin. 16 (4), 577–596.

Zhang, Q., Hu, Y., Zhao, Y., Cheng, J., Fan, W., Hu, D., Shi, F., Cao, S., Zhou, Y., Yang, Y., Liu, X., Zheng, H., Liang, D., Hu, Z., 2024. Deep generalized learning model for PET image reconstruction. IEEE Trans. Med. Imaging 43 (1), 122–134.

Zhou, B., Xie, H., Liu, Q., Chen, X., Guo, X., Feng, Z., Hou, J., Zhou, S.K., Li, B., Rominger, A., Shi, K., Duncan, J.S., Liu, C., 2023. FedFTN: Personalized federated learning with deep feature transformation network for multi-institutional low-count PET denoising. Med. Image Anal. 90, 102993.

Zhu, M., Zhao, M., Yao, M., Guo, R., 2023. A generative adversarial network with zero-shot learning for positron image denoising. Sci. Rep. 13 (1), 1051.